\documentclass[cite,figures,bm]{epl}
\newcommand{\gtrsim}{\stackrel{>}{\sim}}
\title{Broad versus narrow Fano-Feshbach resonances in the BCS-BEC 
crossover with trapped Fermi atoms}
\shorttitle{Broad vs narrow Fano-Feshbach resonances}
\author{S. Simonucci \and P. Pieri \and G.C. Strinati}
\institute{ Dipartimento di Fisica, UdR INFM, 
Universit\`{a} di Camerino, I-62032 Camerino, Italy}


\begin{document}

\maketitle

\begin{abstract}
With reference to the broad and narrow Fano-Feshbach resonances of 
$^{6}$Li at about $822G$ and $543G$, we show 
that for the broad resonance a molecular coupled-channel calculation can be 
mapped with excellent accuracy onto an effective single-channel problem 
with a contact interaction.
This occurs for a wide enough range of the magnetic field, that the full 
BCS-BEC crossover can be realized with a typical trap.
For the narrow resonance, the mapping onto a single-channel problem 
\emph{and} the realization of the BCS-BEC crossover are restricted to too 
narrow a range of the magnetic field to obtain them in practice.
A general criterion is also formulated for deciding whether the BCS-BEC 
crossover can be exhausted within the single-channel model for any specific 
Fano-Feshbach resonance.
In this way, the BCS-BEC crossover for Fermi atoms with the broad resonance is
placed on the same footing as the corresponding crossover  
for different physical systems.
\end{abstract}

Fano-Feshbach (FF) resonances \cite{FF} are currently used to  
control the effective atom-atom interaction in trapped Fermi 
gases \cite{Hulet-2003,Grimm-2004,Ketterle-2004,Jim-2004}, for realizing
 the BCS-BEC crossover \cite{Eagles,Leggett,NSR} from overlapping Cooper pairs 
to non-overlapping composite bosons at low enough temperature.
Study of this crossover was originally motivated by the condensation of 
excitons in solids \cite{crossover-excitons}, and more recently applied 
to nuclei \cite{crossover-nuclei} and high-temperature 
superconductors \cite{crossover-HTS}.

In this context, a large amount of work has been made by adopting a contact 
potential for the effective fermion-fermion attraction,
regularized in terms of the scattering length $a_{F}$
\cite{Randeria-1993,Haussmann-1993,PS-2000}.
The use of this potential considerably simplifies the 
many-body diagrammatic structure, both in the normal 
\cite{Haussmann-1993,PS-2000} and broken-symmetry 
\cite{Haussmann-1993,APS-2003} phases.
By this approach, only fermionic degrees of freedom are retained in the 
many-body Hamiltonian, in the same spirit of the original BCS theory
\cite{Schrieffer}. 
Since just the quantity $a_{F}$ is varied in a controlled way by
sweeping the magnetic field across a FF resonance, it would 
appear that trapped Fermi gases constitute an ideal testing ground for
 many-body theories based on the above two-body interaction.
Previous theoretical work using the same interaction could thus 
be adapted to trapped Fermi gases with limited effort.

The use of the \emph{same} effective two-body interaction for such different 
systems like high-temperature superconductors and trapped 
Fermi gases may also lead to the emergence of \emph{universal} features.
This should be especially desirable, as the insights gleaned from the BCS-BEC 
crossover with trapped Fermi gases could lead to a better understanding of 
high-temperature superconductors \cite{Thomas-AS-2004}.

Most theoretical work on the BCS-BEC crossover with trapped Fermi gases, 
however, has been formulated with a fermion-boson model, aiming at 
incorporating the molecular states coupled in a FF resonance 
\cite{res_superfluid}.
Inclusion of resonance processes in the BCS-BEC crossover was then 
named ``resonance superfluidity".
This was claimed to result in a different BCS-BEC 
crossover \cite{Stoof-2004} from that discussed for high-temperature 
superconductors. 
A fermion-boson model was actually proposed some time ago for the study of 
high-temperature superconductors \cite{Ranninger-1985}.
This model was, however, conceived to include phenomenologically the 
coupling of fermions to a boson mode, which represents fluctuations 
internal to the fermion system (and not a two-fermion state as in resonance 
superfluidity).

Given the current experimental advances on the BCS-BEC crossover with trapped 
Fermi atoms, it seems timely and important to settle the issue of which 
(single- vs multi-channel) model is relevant for an accurate description
of this crossover.
By performing \emph{ab initio\/} calculations of molecular $^{6}$Li in
the presence of a magnetic field, comparison will be made between 
the two FF resonances occurring at (about) $822G$ and $543G$, which we shall 
regard as representative of ``broad'' and ``narrow'' resonances, respectively.
We will show that, for the broad resonance, the outcomes of the 
coupled-channel calculation can be mapped with excellent accuracy onto the 
results of the single-channel model with a contact potential over a wide 
magnetic field range.
On the contrary, for the narrow resonance this mapping can be realized only 
for too narrow a region of the magnetic field to be experimentally accessible.

These two resonances have been selected to pursue the realization of the 
BCS-BEC crossover with trapped Fermi atoms 
(cf. Refs.\cite{Hulet-2003} and \cite{Grimm-2004,Ketterle-2004} for the narrow
and broad resonance, in the order).
We will conclude that the experimental realization of the BCS-BEC crossover 
can be made, in practice, only with the broad resonance and that
the use of the single-channel model is correspondingly appropriate to describe
the crossover\cite{combescot}.
The ultimate reason for the ``broad'' and ``narrow'' nature of the FF 
resonances here considered will also be accounted for.

The key theoretical issue is whether a \emph{minimal} description 
of many-body effects in these systems can be given in 
terms of a single-channel model (where the scattering length $a_{F}$ is the 
only relevant parameter) or of a multi- (possibly two-) channel model (where 
additional parameters, like the effective range of the potential, appear).
In the theory of resonance superfluidity \cite{res_superfluid}, the need for
a multi-channel model to describe 
the BCS-BEC crossover with trapped Fermi atoms was assumed, arguing 
generically that the energy dependence of two-body scattering matters for this
problem \cite{Kokkelmans-2002}.
The point is, however, that the experimental conditions  
introduce the average interparticle distance as an intrinsic length scale, so 
that only after combining this length scale with the two-body properties of a 
given FF resonance one can decide whether the single- or multi-channel model 
is adequate.

\begin{figure}
\twofigures[width=8.2cm]{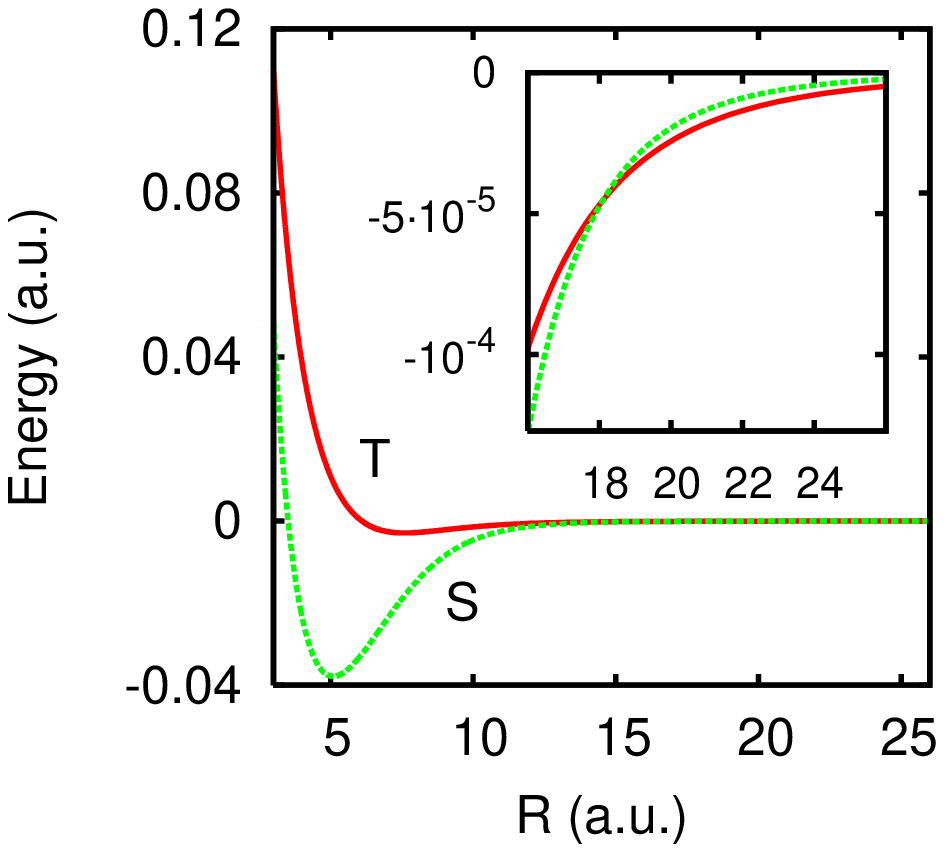}{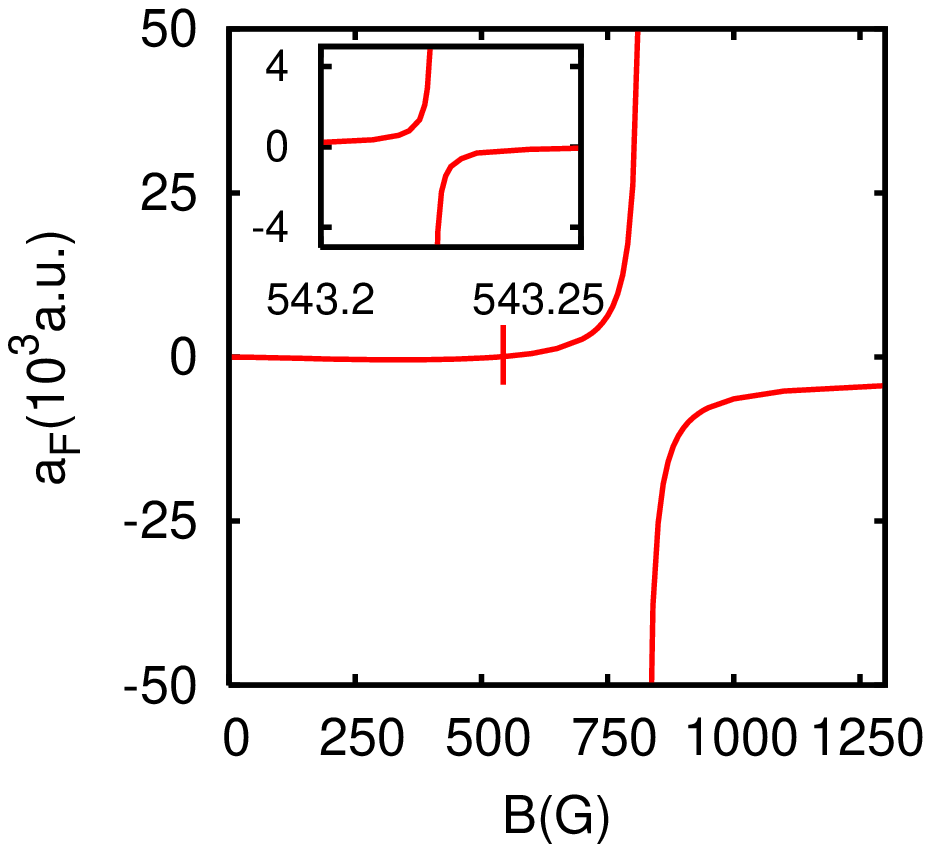}
\caption{Molecular energy curves plus (diagonal) hyperfine and 
Zeeman contributions (taken at $820G$) for the singlet ($S$) 
and triplet ($T$) states of lowest energy vs the nuclear separation $R$. The 
inset shows the details of the crossing at $R=18$ a.u..}
\caption{Scattering length $a_{F}$ vs 
magnetic field $B$.
The inset shows the details of the narrow resonance at about $543G$.}
\end{figure}

The molecular calculation for the biatomic molecule of $^{6}$Li is set up by 
the Born-Oppenheimer scheme.
The auxiliary electronic Schr\"odinger equation at fixed nuclear separation 
$R$ is solved via a standard Configuration Interaction method, with zero  
projection of the total electron angular momentum along the nuclear axis.
The coefficient of the $R^{-6}$ tail of the electronic energy curves is 
fine-tuned (to within a few percent), in order to fix the positions of the FF 
resonances at the experimental values $(822\pm3)G$ 
\cite{Grimm-2004,Ketterle-2004} and $(543.23\pm.05)G$ \cite{Hulet-2003} 
(these positions turn, in fact, out to be quite sensitive to the details of 
the curves for the electronic energy).
[In practice, the $R^{-6}$ tail of these energy curves is set to zero for $R$ 
larger than $R_{0}=4000$ a.u..]
With this slight adjustment only, the scattering length is found to have a 
zero crossing at $B=533G$, which is quite close to the experimental value 
$(528 \pm 4)G$ \cite{Thomas-I-2002}.
In addition, the calculated hyperfine splitting between the atomic $F=3/2$ and
 $F=1/2$ states turns out to be 217 MHz, to be compared with the experimental 
value 228 MHz. 
The nuclear wave equation contains the hyperfine and Zeeman couplings, and is 
solved only for $s$-wave relative motion.
The low-energy sector relevant to the two FF resonances is thus 
spanned by five channels, corresponding to the singlet $^{1}\Sigma_{g}^{+}$
with spin configurations $|0,0\rangle_{{\rm e}} |0,0\rangle_{{\rm n}}$ and 
$|0,0\rangle_{{\rm e}}
|2,0\rangle_{{\rm n}}$ 
(which will be referred to as the ``resonance'' channels), and to the triplet 
$^{3}\Sigma_{u}^{+}$ with spin configurations $|1,1\rangle_{{\rm e}} 
|1,\bar{1}\rangle_{{\rm n}}$, 
$|1,0\rangle_{{\rm e}} |1,0\rangle_{{\rm n}}$, and 
$|1,\bar{1}\rangle_{{\rm e}} |1,1\rangle_{{\rm n}}$ (which will be 
referred to as the ``scattering'' channels), for the electronic (${\rm e}$) 
and nuclear (${\rm n}$) spin functions with $F_{z}=0$.

Figures 1 and 2 show representative outcomes of our molecular calculation.
In particular, in Fig.1 the (diagonal) hyperfine and Zeeman contributions 
(taken at $820G$) are added to the electronic energy curves of lowest 
energy for the singlet ($S$) and triplet ($T$).
Note in the inset the crossing of the two curves induced by the Zeeman 
coupling, which makes the lowest energy threshold of triplet character.
In Fig.2 the calculated scattering length $a_{F}$ is shown 
from $B=0$ to $B=1300G$.
The inset shows the details of $a_F$ for the narrow resonance at about $543G$.

Several quantities of interest can be extracted from this calculation.
Below resonance, we compare:
(i) The binding energy $E_{{\rm b}}$ (obtained from the multi-channel calculation) to
the value $\epsilon_{0} = (M a_{F}^{2})^{-1}$ for the contact potential, where
$M$ is the nuclear mass;
(ii) The corresponding (root-mean-square) radius $\bar{R}$ of the full bound 
wave function to the value $a_{F}/\sqrt{2}$ for the contact potential.
We also calculate:
(iii) The (root-mean-square) radius $\bar{R}_{2}$ associated with the 
component of the total wave function in the resonance channel;
(iv) The (squared) projection $|w_{1}|^{2}$ of the total wave function in the 
scattering channel;
(v) The corresponding projection $\langle \phi_{{\rm cp}}|w_{1}\rangle$ onto 
the bound wave function $\phi_{{\rm cp}}=(\sqrt{2/a_{F}}) \exp (-R/a_{F})$ 
for the contact  potential.

\begin{table}
\caption{Comparison of the molecular calculation with the 
effective single-channel model, for the broad resonance at about $822G$ 
and for the narrow resonance at about $543G$.}
\begin{largetabular}{cccccccc}
B (G) & $a_{F} (10^3$a.u.)&$E_{\rm{b}}/\epsilon_0$& $\sqrt{2}\bar{R}/a_F$ & $\bar{R}_2/\bar{R}$ 
& $|w_1|^2$ & $\langle \phi_{{\rm cp}}| w_1 \rangle$& 
$r_0 (10^3$a.u.)\\
\hline
650& 1.29 & 1.068 & 1.00053& .039 &.99669
& .972 & .085 \\
750&6.26& 1.014 &1.00004 &.008  & .99986 
& .994& .087 \\
800& 26.2 & 1.003 &1.00024  &.002  &.99998
& .999  & .088 \\
850& -25.3& & & &
& &  .088 \\
1100& -5.17 & & & & & & .090 \\
1300&-4.35 & & & &
& & .090 \\
\hline
543.2200& 2.09& .0425 & 1.148  &.022 & .062 & .194
&-121  \\
543.2210& 4.85 & .0581 & 1.517 &.007& .134 & .289&
-121 \\
543.2216 & 41.7 & .9066 &.5853 &.002& .304  & .550  &-124 \\
543.2218 &-34.3 & & & & & &-125 \\
543.2220 &-10.7 & & & & & &-127 \\
543.2225 &-4.22 & & & & & &-126\\
\end{largetabular}
\end{table}
                                   
Table I reports all these quantities for the two resonances.
Note how, for the broad resonance, the effective single-channel 
model with a contact potential reproduces with excellent accuracy all 
results obtained by the multi-channel calculation.
Note, in particular, how the bound-state wave function of the multi-channel 
calculation projects almost completely in the scattering channel,
which is the sole considered by the effective single-channel model.
The spatial extension $\bar{R}_2$ of the boson introduced in  
resonance-superfluidity theory \cite{res_superfluid} remains instead quite 
smaller than the extension $\bar{R}$ of the true molecular wave function 
(which corresponds to the internal wave function of the composite boson).
A different situation occurs for the narrow resonance.
In this case, the effective single-channel model is appropriate only too 
close to the resonance for being of practical relevance.

To study the BCS-BEC crossover with FF resonances, the above analysis of the 
molecular calculation must be complemented by the value of the Fermi wave 
vector $k_{F}$ (determined by the total number of 
Fermi atoms in the trap and the trap frequencies) and by the minimum 
experimental accuracy for the magnetic field.

Irrespective of the underlying theoretical model, the dimensionless parameter 
$(k_{F} a_{F})^{-1}$ should exhaust the BCS-BEC crossover within a range 
$\approx 1$ about the unitarity limit at $(k_{F} a_{F})^{-1}=0$. 
This is because the BCS (with $a_{F} < 0$) and BEC (with $a_{F} > 0$) 
regimes should be reached when $k_{F} |a_{F}|\ll 1$, while in the crossover 
region $k_{F} |a_{F}|$ diverges.
To span the crossover, one thus needs to identify (at least) three 
representative values (say, -1.0, 0.0, and 1.0) of  
$(k_{F} a_{F})^{-1}$, by tuning the magnetic field that controls the FF 
resonance. 
With $k_{F}= 2 \div 3 \times 10^{-4}$ a.u. for the experiments of 
Refs.~\cite{Grimm-2004} and \cite{Ketterle-2004}, these values of 
$(k_{F} a_{F})^{-1}$ correspond approximatively to the magnetic field values 
$(1300,822,730)G$, which are separated by a 
step $\delta B \gtrsim 100G$ much larger than the minimum experimental 
accuracy.
For these values of $B$, the single-channel model is totally appropriate to 
describe the two-body scattering, as seen from Table I. 
The situation is reversed for the narrow resonance.
In this case, the above values of $(k_{F} a_{F})^{-1}$ are realized by setting
the magnetic field at $(543.2225,543.2217,543.2209)G$,  
with a step $\delta B \simeq 0.001G$ fifty times smaller than 
the minimum experimental accuracy \cite{Hulet-2003}.

To complete the mapping into the single-channel model, there remains 
to identify the effective spin states $|\!\uparrow\rangle_{\rm{eff}}$ and 
$|\!\downarrow\rangle_{\rm{eff}}$.
From the molecular calculation, the wave function for the broad resonance 
is found to be (essentially) a triplet both for electrons and nuclei.
This implies the identification
$|\!\uparrow\rangle_{\rm{eff}} \leftrightarrow |\frac{1}{2}, 
-\frac{1}{2}\rangle_{\rm{e}} \, |1,0\rangle_{\rm{n}}$ and    
$|\!\downarrow\rangle_{\rm{eff}} \leftrightarrow |\frac{1}{2}, 
-\frac{1}{2}\rangle_{\rm{e}} \, |1,1\rangle_{\rm{n}}$
for the separate atoms, as well as the effective singlet configuration 
$(|\!\uparrow\rangle_{\rm{eff}}^{(1)} \, 
|\!\downarrow\rangle_{\rm{eff}}^{(2)}  - |\!\uparrow\rangle_{\rm{eff}}^{(2)} 
\, |\!\downarrow\rangle_{\rm{eff}}^{(1)})/\sqrt{2}$
for the pair of atoms (with the labels $1$ and $2$ referring to the two 
atoms).

Further information can be extracted from the molecular calculation when the 
total energy is above threshold.
It has been asserted \cite{res_superfluid} that, 
in resonance conditions, the scattering length $a_{F}$ no longer suffices to 
account for the scattering properties, since these depend 
strongly on energy.
Quite generally, the scattering amplitude takes the form 
$f(k) = (g(k) - i k)^{-1}$, where the wave vector $k$ is associated with 
the kinetic energy above threshold. 
For a contact potential $g(k)=-a_{F}^{-1}$, while for a multi-channel problem 
$g(k)$ in the scattering channel contains additional $k$-dependent terms.
In particular, at the lowest order in $k$, $g(k)=-a_{F}^{-1}+r_{0}k^{2}/2$ 
where $r_{0}$ identifies the ``effective range'' 
of the potential. 
Near resonance (when $a_{F}^{-1} \approx 0$), $f(k)$ has a strong 
$k$-dependence even in the absence of $r_{0}$.
The term $r_{0}k^{2}/2$ begins to matter only when 
$r_{0} k \gtrsim 1$.
For the BCS-BEC crossover, a natural cutoff for $k$ is provided by $k_{F}$ in 
the weak-to-intermediate coupling regime and by $a_{F}^{-1}$ in the 
strong-coupling regime.
The term $r_{0}k^{2}/2$ is thus {\em irrelevant} provided 
$|r_{0}| k_{F} \ll 1$ and $|r_{0}| \ll |a_{F}|$.
The values of $r_{0}$ are reported in Table I for the two 
resonances.
For the broad resonance, $r_{0}$ is positive (as it is the case for a 
potential problem) and remains much smaller than $|a_{F}|$.
For the narrow resonance, none of these properties are verified. 
Since for the broad resonance the product $r_{0} k_{F}$ is smaller than 
$10^{-2}$ (with a typical value $k_{F}\approx 10^{-4}$ a.u.), we conclude 
that the energy dependence of the scattering properties ({\em over and 
above\/} that resulting from a contact potential) is actually irrelevant when 
realizing the BCS-BEC crossover with the broad resonance. 
\begin{figure}
\onefigure[width=8.6cm]{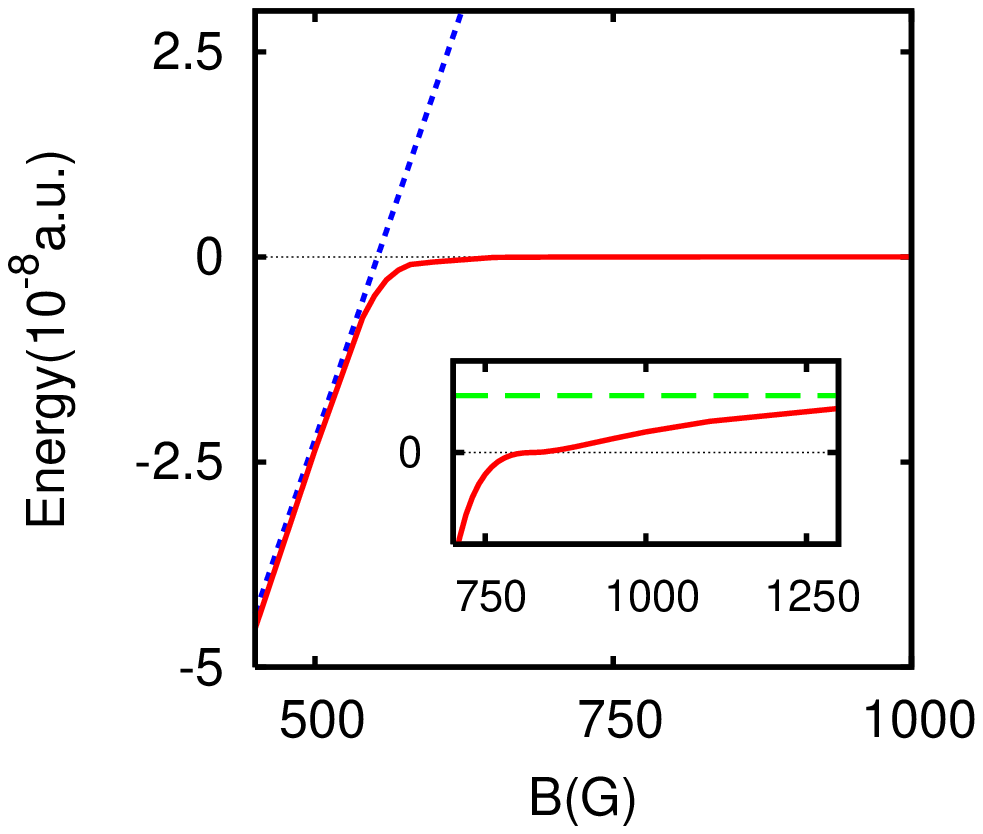}
\caption{Mechanism for the occurrence of the narrow and broad FF resonances 
(see text).}
\label{mechanism}
\end{figure}

It is, finally, of interest to understand the reason why the 
resonances here considered are broad and narrow relative to each other.
Out of the five channels spanning the low-energy sector, the {\em two\/} 
$^{1}\Sigma_{g}^{+}$ singlets with spin configurations 
$|0,0>_{\rm{e}} |0,0>_{\rm{n}}$ and  $|0,0>_{\rm{e}} |2,0>_{\rm{n}}$ 
interact mostly with the $^{3}\Sigma_{u}^{+}$ triplet with spin configuration
 $|1,\bar{1}>_{\rm{e}} |1,1>_{\rm{n}}$.
In Fig.~3 the magnetic field dependence of the energy of the last singlet 
bound state (with $38$ nodes) is plotted (dashed line) relative to the 
threshold of the triplet channel (dotted line) for the case when these 
channels are completely decoupled.
These curves cross each other at (about) $550G$.
The triplet channel, in addition, is found to have a virtual state at 
threshold \cite{virtual-state}, as the large negative value of the background 
scattering length $a_{\rm{bg}}$ indicates ($a_{\rm{bg}}\simeq-3790$ a.u. when 
calculated with this channel only).
[This value results from our {\em ab initio\/} calculation, 
where the only slightly adjusted parameters are the coefficients of the 
$R^{-6}$ tail of the electronic energy curves. It differs by about 
40\% from that reported in the literature~\cite{Thomas-I-2002}, where a larger 
number of adjustable
parameters is however used. This difference in 
$a_{\rm{bg}}$ does not affect appreciably the values in Table I but for the 
function $a_F(B)$.]

Out of the two singlet channels, it is always possible to find a linear 
combination that decouples from the triplet virtual state, thus
crossing threshold at (about) $550G$.
The other combination which couples with the triplet is forced, 
correspondingly, to have an \emph{avoided crossing} when the interaction 
between singlets and 
triplet is restored (full line).
The combination which decouples from the triplet results thus in a ``narrow'' 
resonance (with a small broadening provided by the coupling to the 
rest of the continuum states), and mantains its singlet 
character through the crossing at (about) $550G$. 
The combination which couples with the triplet, on the other hand, acquires 
(almost) full triplet character past the avoided crossing.
For large magnetic field, the energy of the latter tends asymptotically to the
broadening $(M a_{{\rm bg}}^2)^{-1}$ of the virtual state (long-dashed line in
the inset of Fig.~3), thus crossing 
threshold quite slowly.
The slow convergence to the asymptote eventually accounts for the ``broad'' 
nature of the resonance vs the magnetic field.

In conclusion, by extracting the relevant information from a coupled-channel 
calculation for the broad and narrow FF resonances of $^{6}$Li, 
a single-channel model with a contact interaction proves
adequate to describe the BCS-BEC crossover with the broad resonance.
In this way, a connection is established between the BCS-BEC crossover 
with trapped Fermi atoms and analogous crossovers with other systems.
Trapped Fermi atoms are thus ideally suited for studying 
the BCS-BEC crossover in a universal fashion, irrespective of  
mechanisms specific to these simple systems.

\acknowledgments
This work was partially supported by the Italian MIUR with contract Cofin-2003
 ``Complex Systems and Many-Body Problems''.


\begin{thebibliography}{99}

\bibitem{FF} 
\Name{Fano U.}\REVIEW{Nuovo Cimento}{12}{1935}{156} and 
\REVIEW{Phys. Rev.}{124}{1961}{1866};
 \Name{Feshbach H.}
 \REVIEW{Ann. Phys.}{19}{1962}{287}.
             
\bibitem{Hulet-2003}
\Name{Strecker K.E., Partridge G.B. \and Hulet R.G.} 
\REVIEW{Phys. Rev. Lett.}{91}{2003}{080406}.

\bibitem{Grimm-2004}
\Name{Bartenstein M., Altmeyer A., Riedl S., Jochim S., 
Chin C., Hecker Denschlag J. \and Grimm R.}
\REVIEW{Phys. Rev. Lett.}{92}{2004}{120401}.

\bibitem{Ketterle-2004}
\Name{Zwierlein M.W., Stan C.A., Schunck C.H., Raupach S.M.F., Kerman  A.J. 
\and Ketterle  W.}
\REVIEW{Phys. Rev. Lett.}{92}{2004}{120403}.

\bibitem{Jim-2004}
\Name{Regal C.A., Greiner M. \and Jin D.S.}
\REVIEW{Phys. Rev. Lett.} 
{92}{2004}{040403}. 

\bibitem{Eagles}
\Name{Eagles D.M.}
\REVIEW{Phys. Rev.}{186}{1969}{456}.

\bibitem{Leggett}\Name{Leggett A.J.}
\Book{Modern Trends in the Theory of 
Condensed Matter}
\Editor{Pekalski A. \and Przystawa R.},
Lecture Notes in Physics
\Vol{115}
\Publ{Springer, Berlin}
\Year{1980}
\Page{13}.

\bibitem{NSR}
\Name{Nozi\`{e}res P. \and Schmitt-Rink S.}
\REVIEW{J. Low. Temp. Phys.}{59}{1985}{195}.

\bibitem{crossover-excitons}
\Name{Lai C.W., Zoch J., Gossard A.C. \and Chemla D.S.}
\REVIEW{Science}{303}{2004}{503}, and references quoted therein.

\bibitem{crossover-nuclei}
\Name{Baldo M., Lombardo U., \and Schuck P.}
\REVIEW{Phys. Rev. C}{52}{1995}{975}.

\bibitem{crossover-HTS} 
\Name{Loktev V.M., Quick R.M. \and Sharapov S.G.}
\REVIEW{Phys. Rep.}{349}{2001}{1}, and references quoted therein.

\bibitem{Randeria-1993}
\Name{S\'a de Melo C.A.R., Randeria  M. \and  Engelbrecht J.R.}
\REVIEW{Phys. Rev. Lett.}{71}{1993}{3202}.

\bibitem{Haussmann-1993}
\Name{Haussmann R.}
\REVIEW{Z. Phys. B}{91}{1993}{291}.

\bibitem{PS-2000}
\Name{Pieri P. \and Strinati G.C.}
\REVIEW{Phys. Rev. B}{61}{2000}{15370}.

\bibitem{APS-2003}
\Name{Andrenacci N., Pieri P. \and Strinati G.C.}
\REVIEW{Phys. Rev. B}{68}{2003}{144507}.

\bibitem{Schrieffer}
\Name{Schrieffer J.R.} 
\Book{Theory of Superconductivity} 
\Publ{Benjamin, New York} 
\Year{1964}.

\bibitem{Thomas-AS-2004} 
\Name{Thomas J.E. \and Gehm  M.E.}
\REVIEW{American Scientist}{92}{2004}{238}.

\bibitem{res_superfluid}
\Name{Timmermans  E., Furuya  K., Milonni P.W. \and Kerman A.K.}
\REVIEW{Phys. Lett. A}{285}{2001}{228}; 
\Name{Holland M.,Kokkelmans S.J.J.M.F., Chiofalo  M.L., \and Walser  R.}
\REVIEW{Phys. Rev. Lett.}{87}{2001}{120406};
\Name{Ohashi Y. \and Griffin  A.}
\REVIEW{Phys. Rev. Lett.}{89}{2002}{130402}.

\bibitem{Stoof-2004}
\Name{Falco G.M. \and Stoof H.T.C.}
\REVIEW{Phys. Rev. Lett.}{92}{2004}{130401}.

\bibitem{Ranninger-1985}
\Name{Ranninger J. \and Robaszkiewicz S.} 
\REVIEW{Physica B}{135}{1985}{468}.

\bibitem{combescot} 
\Name{Combescot R.}
\REVIEW{Phys. Rev. Lett.}{91}{2003}{120401} 

\bibitem{Kokkelmans-2002}
\Name{Kokkelmans S.J.J.M.F., Milstein  J.N., Chiofalo  M.L., Walser R. \and 
Holland M.J.}
\REVIEW{Phys. Rev. A}{65}{2002}{053617}.


\bibitem{Thomas-I-2002} 
\Name{O'Hara K.M., Hemmer S.L., Granade S.R., Gehm M.E., Thomas J.E., 
Venturi V., Tiesinga E. \and Williams C.J.} 
\REVIEW{Phys. Rev. A}{66}{2002}{041401(R)}.

                     
\bibitem{virtual-state} The relevance of resonances in the scattering 
channel was discussed by 
\Name{Marcelis B., van Kempen E.G.M., Verhaar B.J. 
\and Kokkelmans S.J.J.M.F.}
\REVIEW{Phys. Rev. A}{70}{2004}{012701}.

\end{thebibliography}
\end{document}